\documentclass[%
 reprint,
superscriptaddress,
longbibliography,
 amsmath,amssymb,
 aps,
]{revtex4-1}

\usepackage{placeins}
\usepackage{graphicx}
\usepackage{dcolumn}
\usepackage{bm}

\begin{document}

\title{Theoretical basis of the diagonal scan method for determining the laser ablation threshold for femtosecond vortex pulses}
\author{Reece N. Oosterbeek}
\email{reece.oosterbeek@auckland.ac.nz}
\affiliation{Photon Factory, The University of Auckland, Auckland, New Zealand}
\affiliation{School of Chemical Sciences, The University of Auckland, Auckland, New Zealand}
\affiliation{The Dodd Walls Centre for Quantum and Photonic Technologies, and The MacDiarmid Institute for Advanced Materials and Nanotechnology, New Zealand}

\author{Simon Ashforth}
\affiliation{Photon Factory, The University of Auckland, Auckland, New Zealand}
\affiliation{The Dodd Walls Centre for Quantum and Photonic Technologies, and The MacDiarmid Institute for Advanced Materials and Nanotechnology, New Zealand}
\affiliation{Department of Physics, The University of Auckland, Auckland, New Zealand}

\author{Owen Bodley}
\affiliation{Photon Factory, The University of Auckland, Auckland, New Zealand}
\affiliation{School of Chemical Sciences, The University of Auckland, Auckland, New Zealand}
\affiliation{The Dodd Walls Centre for Quantum and Photonic Technologies, and The MacDiarmid Institute for Advanced Materials and Nanotechnology, New Zealand}

\author{M. Cather Simpson}
\email{c.simpson@auckland.ac.nz}
\affiliation{Photon Factory, The University of Auckland, Auckland, New Zealand}
\affiliation{School of Chemical Sciences, The University of Auckland, Auckland, New Zealand}
\affiliation{The Dodd Walls Centre for Quantum and Photonic Technologies, and The MacDiarmid Institute for Advanced Materials and Nanotechnology, New Zealand}
\affiliation{Department of Physics, The University of Auckland, Auckland, New Zealand}

\date{\today}

\begin{abstract}
In femtosecond laser micromachining, the ablation threshold is a key processing parameter that characterises the energy density required to cause ablation. Current techniques for measuring the ablation threshold such as the diameter regression and diagonal scan methods are based on the assumption of a Gaussian spatial profile, however no techniques currently exist for measuring the ablation threshold using a non-Gaussian beam shape.
\\
Here we present a formalism of the diagonal scan method for determining the ablation threshold and pulse superposition for femtosecond vortex pulses. To the authors' knowledge this is the first ablation threshold technique developed for pulses with non-Gaussian spatial profiles.
\\
Using this method, the ablation threshold can be calculated using measurement of a single feature (the maximum damage radius $\rho_{max}$), which allows investigations of ablation threshold and incubation effects to be carried out quickly and easily. Extending this method to non-Gaussian beams will allow exploration of new avenues of research, enabling characterisation of the ablation threshold and incubation behaviour for a material when ablated with femtosecond vortex pulses.
\end{abstract}

\maketitle

\section{Introduction}
Femtosecond laser ablation is an advanced materials processing technique that enables microstructures to be fabricated in almost any material with very high accuracy and resolution \cite{Krueger04, Cheng13}. The ultrashort pulse duration leads to non-linear absorption, allowing materials to be ablated regardless of their linear absorption characteristics \cite{Perry99}. In addition, this ultrashort timescale limits energy transfer into the atomic lattice, greatly reducing heat effects, allowing materials to be ablated with little to no damage in the surrounding area \cite{Krueger97}. These advantages make femtosecond laser ablation an attractive prospect for industrial applications.
\\
In all work involving femtosecond laser ablation, the ablation threshold ($F_{th}$, in $J/cm^{2}$) is a key parameter for characterising the interaction between laser and material, and is defined as the minimum energy density required to cause material removal. Therefore, it is of utmost importance that robust and useful methods of measuring the ablation threshold are available and widely applicable. Current methods for determining the ablation threshold are the diameter regression method \cite{Sanner09} and the diagonal scan method \cite{Samad06}. Both of these methods rely on the assumption that ablation is carried out using a laser beam with Gaussian spatial distribution. With rapid advances being made in the area of spatial beam shaping however, this assumption is not always valid. In particular, optical vortex beams have been investigated recently, indicating generation of different nanostructures to those obtained when using a Gaussian beam \cite{Hnatovsky10, Hnatovsky12, Anoop14a, Anoop14b}.
\\
In this work we present a formalism of the diagonal scan method for measuring the femtosecond laser ablation threshold that is applicable to vortex beams. We derive an expression for calculating the ablation threshold based on measurement of the maximum damage radius, and also demonstrate a method for calculating the pulse superposition obtained during a diagonal scan experiment. 

\section{Damage Radius}\label{sec:Radius}
\begin{figure}
	\centering
	\includegraphics[width=1\columnwidth]{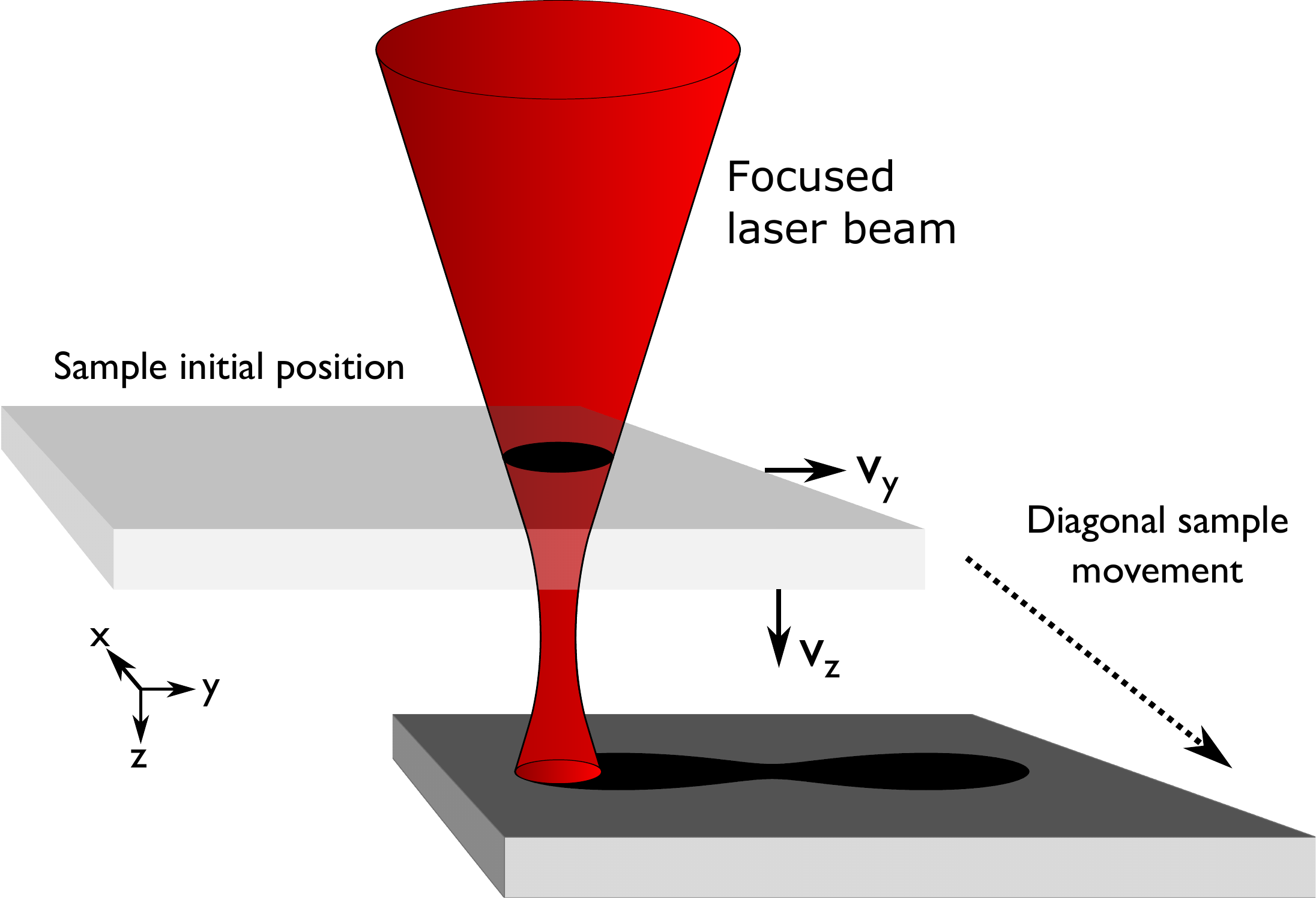}
	\textbf{\caption{\label{fig:Fig_1}Diagram demonstrating the diagonal scan experiment and how the ablation feature is formed.}}
\end{figure}

\begin{figure}
	\centering
	\includegraphics[width=1\columnwidth]{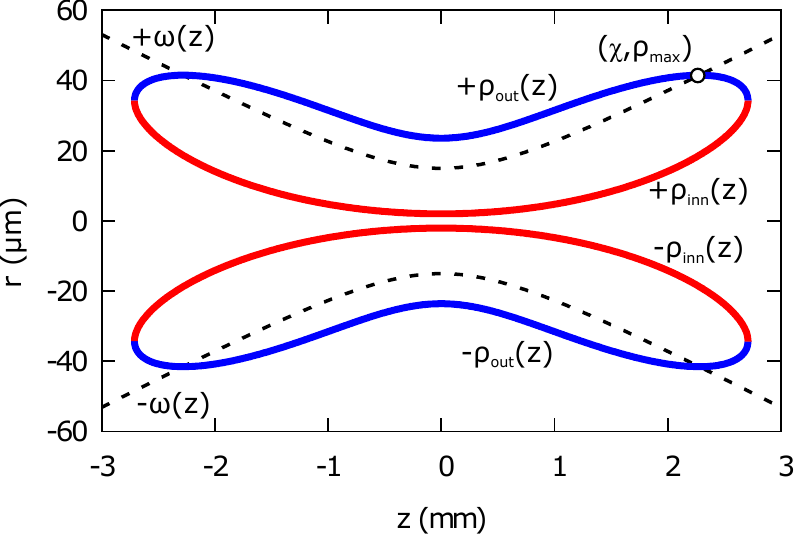}
	\textbf{\caption{\label{fig:Fig_2}Graph showing the damage radius for a vortex beam ($l = 1$), where the inner damage radius (calculated using the principal branch of the Lambert Omega function $W_{0}$) is shown in red, and the outer damage radius (calculated using the non-principal branch of the Lambert Omega function $W_{-1}$)  is shown in blue. The Gaussian beam radius is also shown in black.}}
\end{figure}

\FloatBarrier
To determine an expression for the ablation threshold, we consider a diagonal scan experiment where a sample is translated diagonally through the focal point of a focused laser beam, as shown in Figure \ref{fig:Fig_1}.
\\
The spatial fluence distribution of an optical vortex beam is given by \cite{Hnatovsky10}:
\begin{equation} \label{eq:1}
F\left[\frac{J}{cm^{2}}\right] = \frac{2^{|l|+1}r^{2|l|}e^{\frac{-2r^{2}}{w(z)^{2}}}}{|l|!\pi w(z)^{2(|l|+1)}}E_{0}\left[J\right]
\end{equation}
\\
where: $l = 0, \pm 1, \pm 2, \pm 3.....$ is the topological charge, $r$ is the radial direction (in $cm$), and $w(z)$ is the radius of a Gaussian beam (in $cm$) at $l = 0$, which is equal to:
\begin{equation} \label{eq:2}
w(z) = w_{0}\left| 1 + \frac{iz\lambda}{\pi w_{0}^{2}}\right|
\end{equation}
\\
where $z$ is the propogation direction, $\lambda$ is the wavelength, and $w_{0}$ is the beam waist (all in $cm$).
\begin{equation} \label{eq:3}
w(z) = w_{0}\sqrt{1 + \frac{z^{2}\lambda^{2}}{\pi^{2} w_{0}^{4}}}
\end{equation}
\\
At $l = 0$ Equation \ref{eq:1} simplifies to the equation for a Gaussian beam:
\begin{equation} \label{eq:4}
F\left[\frac{J}{cm^{2}}\right] = \frac{2e^{\frac{-2r^{2}}{w(z)^{2}}}}{\pi w(z)^{2}}E_{0}\left[J\right]
\end{equation}
\\
We can set the damage radius $\rho (z)$ to be equal to the radius at which the damage threshold $F_{th}$ is exceeded:
\begin{equation} \label{eq:5}
F_{th} = \frac{2^{|l|+1}\rho (z)^{2|l|}e^{\frac{-2\rho (z)^{2}}{w(z)^{2}}}}{|l|!\pi w(z)^{2(|l|+1)}}E_{0}
\end{equation}
\begin{equation} \label{eq:6}
\frac{F_{th}|l|!\pi w(z)^{2(|l|+1)}}{2^{|l|+1}E_{0}} = \rho (z)^{2|l|}e^{\frac{-2\rho (z)^{2}}{w(z)^{2}}}
\end{equation}
\begin{equation} \label{eq:7}
\left(\frac{F_{th}|l|!\pi w(z)^{2(|l|+1)}}{2^{|l|+1}E_{0}}\right)^{\frac{1}{|l|}} = \rho (z)^{2}e^{\frac{-2\rho (z)^{2}}{|l|w(z)^{2}}}
\end{equation}
\\
We can set:
\begin{equation} \label{eq:8}
A = \left(\frac{F_{th}|l|!\pi w(z)^{2(|l|+1)}}{2^{|l|+1}E_{0}}\right)^{\frac{1}{|l|}}
\end{equation}
\begin{equation} \label{eq:9}
B = \frac{-2}{|l|w(z)^{2}}
\end{equation}
\\
such that:
\begin{equation} \label{eq:10}
A = \rho (z)^{2}e^{B\rho (z)^{2}}
\end{equation}
\begin{equation} \label{eq:11}
AB = B\rho (z)^{2}e^{B\rho (z)^{2}}
\end{equation}
\\
This is an equation of the form $z = ue^{u}$, which can be solved according to the Lambert Omega function \cite{Corless96} by $u = W(z)$.
\begin{equation} \label{eq:12}
B\rho (z)^{2} = W(AB)
\end{equation}
\begin{equation} \label{eq:13}
\rho (z) = \sqrt\frac{W(AB)}{B}
\end{equation}
\\
The Lambert Omega function is multivalued (except at zero), therefore for any $z$ two radii can be defined, the inner $(\rho_{inn})$ and outer $(\rho_{out})$ damage radius, which can be calculated using the principal $(W_{0})$ and non-principal $(W_{-1})$ branches of the Lambert Omega function:
\begin{equation} \label{eq:144}
\rho _{inn}(z) = \sqrt\frac{W_{0}(AB)}{B}
\end{equation}
\begin{equation} \label{eq:145}
\rho _{out}(z) = \sqrt\frac{W_{-1}(AB)}{B}
\end{equation}
Equations \ref{eq:144} and \ref{eq:145}  describe the damage radius as a function of the propogation direction $z$, and are shown in Figure \ref{fig:Fig_2}. For application in a diagonal scan ablation threshold measurement technique, we are only interested in the outer damage radius $(\rho_{out})$. For this reason we do not consider the inner damage radius $(\rho_{inn})$ further, and all references to $\rho$ refer to the outer damage radius $(\rho_{out})$.
\\
For real numbers, the non-principal branch of the Lambert Omega function has the limits:
\begin{equation} \label{eq:61}
u = W_{-1}(x) \text{ for } 0 \geq x \geq \frac{-1}{e}
\end{equation}
\\
The implications of this limit will be discussed further in Section \ref{sec:Omega}.

\section{Finding the Maxima}\label{sec:Max}

To find $\rho_{max}$ we must set $\frac{d\rho}{dz} = 0$. Let:
\begin{equation} \label{eq:14}
\rho (z) = \sqrt\frac{W_{-1}(AB)}{B} = \sqrt{C}
\end{equation}
\begin{equation} \label{eq:15}
C = \frac{D}{B}
\end{equation}
\begin{equation} \label{eq:16}
D = W_{-1}(G)
\end{equation}
\begin{equation} \label{eq:17}
G = AB = \frac{(F_{th}|l|!\pi )^{\frac{1}{|l|}}w_{0}^{2(1+\frac{1}{|l|})}H^{1+\frac{1}{|l|}}}{2^{1+\frac{1}{|l|}}E_{0}^{\frac{1}{|l|}}} \times \frac{-2}{|l|w_{0}^{2}H}
\end{equation}
\begin{equation} \label{eq:18}
G = \frac{-(F_{th}|l|!\pi )^{\frac{1}{|l|}}w_{0}^{\frac{2}{|l|}}H^{\frac{1}{|l|}}}{2^{\frac{1}{|l|}}E_{0}^{\frac{1}{|l|}}|l|}
\end{equation}
\begin{equation} \label{eq:19}
H = \frac{z^{2}\lambda^{2}}{\pi^{2}w_{0}^{4}} + 1
\end{equation}
\begin{equation} \label{eq:20}
B = \frac{-2}{|l|w_{0}^{2}H}
\end{equation}
\\
By applying the chain rule to these definitions, we can obtain an expression for  $\frac{d\rho}{dz}$
\begin{equation} \label{eq:21}
\frac{d\rho}{dz} = \frac{d\rho}{dC} \times \frac{dC}{dz}
\end{equation}
\begin{equation} \label{eq:22}
\frac{dC}{dz} = \frac{\frac{dD}{dz}B - D\frac{dB}{dz}}{B^{2}} 
\end{equation}
\begin{equation} \label{eq:23}
\frac{dD}{dz} = \frac{dD}{dG} \times \frac{dG}{dH} \times \frac{dH}{dz}
\end{equation}
\begin{equation} \label{eq:24}
\frac{dB}{dz} = \frac{dB}{dH} \times \frac{dH}{dz}
\end{equation}
\\
Defining these derivatives:
\begin{equation} \label{eq:25}
\frac{dH}{dz} = \frac{2z\lambda^{2}}{\pi^{2}w_{0}^{4}}
\end{equation}
\begin{equation} \label{eq:26}
\frac{dB}{dH} = \frac{2}{|l|w_{0}^{2}H^{2}} = \frac{2}{|l|w_{0}^{2}(\frac{z^{2}\lambda^{2}}{\pi^{2}w_{0}^{4}} + 1)^{2}}
\end{equation}
\begin{equation} \label{eq:27}
\frac{dB}{dH} = \frac{2}{|l|w_{0}^{2}(\frac{z^{4}\lambda^{4}}{\pi^{4}w_{0}^{8}} + \frac{2z^{2}\lambda^{2}}{\pi^{2}w_{0}^{4}} + 1)}
\end{equation}
\begin{equation} \label{eq:28}
\frac{dG}{dH} = \frac{-(F_{th}|l|!\pi )^{\frac{1}{|l|}}w_{0}^{\frac{2}{|l|}}H^{\frac{1}{|l|}-1}}{2^{\frac{1}{|l|}}E_{0}^{\frac{1}{|l|}}|l|^{2}}
\end{equation}
\begin{equation} \label{eq:29}
\frac{d\rho}{dC} = \frac{1}{2\sqrt{C}}
\end{equation}
\begin{equation} \label{eq:30}
\frac{dD}{dG} = \frac{W_{-1}(G)}{G(1+W_{-1}(G))}
\end{equation}
\\
We now evaluate the derivatives. From equations \ref{eq:24}, \ref{eq:25} and \ref{eq:26}:
\begin{equation} \label{eq:31}
\frac{dB}{dz} = \frac{dB}{dH} \times \frac{dH}{dz} = \frac{2}{|l|w_{0}^{2}(\frac{z^{4}\lambda^{4}}{\pi^{4}w_{0}^{8}} + \frac{2z^{2}\lambda^{2}}{\pi^{2}w_{0}^{4}} + 1)} \times \frac{2z\lambda^{2}}{\pi^{2}w_{0}^{4}}
\end{equation}
\begin{equation} \label{eq:32}
\frac{dB}{dz} = \frac{4z\lambda^{2}}{|l|\pi^{2}w_{0}^{6}(\frac{z^{4}\lambda^{4}}{\pi^{4}w_{0}^{8}} + \frac{2z^{2}\lambda^{2}}{\pi^{2}w_{0}^{4}} + 1)}
\end{equation}
\begin{equation} \label{eq:33}
\frac{dB}{dz} = \frac{4z\lambda^{2}}{|l|(\frac{z^{4}\lambda^{4}}{\pi^{2}w_{0}^{2}} + 2z^{2}\lambda^{2}w_{0}^{2} + \pi^{2}w_{0}^{6})}
\end{equation}
\\
Let: 
\begin{equation} \label{eq:34}
I = \frac{z^{4}\lambda^{4}}{\pi^{2}w_{0}^{2}} + 2z^{2}\lambda^{2}w_{0}^{2} + \pi^{2}w_{0}^{6}
\end{equation}
\begin{equation} \label{eq:35}
\frac{dB}{dz} = \frac{4z\lambda^{2}}{|l|I}
\end{equation}
\\
From equations \ref{eq:23}, \ref{eq:25}, \ref{eq:28} and \ref{eq:30}:
\begin{equation} \label{eq:36}
\frac{dD}{dz} = \frac{dD}{dG} \times \frac{dG}{dH} \times \frac{dH}{dz}
\end{equation}
\begin{multline} \label{eq:37}
	\frac{dD}{dz} = \frac{W_{-1}(G)}{G(1+W_{-1}(G))} \times \\ \frac{-(F_{th}|l|!\pi )^{\frac{1}{|l|}}w_{0}^{\frac{2}{|l|}}H^{\frac{1}{|l|}-1}}{2^{\frac{1}{|l|}}E_{0}^{\frac{1}{|l|}}|l|^{2}} \times \frac{2z\lambda^{2}}{\pi^{2}w_{0}^{4}}
\end{multline}
\begin{multline} \label{eq:38}
	\frac{dD}{dz} = \frac{W_{-1}(G)}{1+W_{-1}(G)} \times \frac{2^{\frac{1}{|l|}}E_{0}^{\frac{1}{|l|}}|l|}{-(F_{th}|l|!\pi )^{\frac{1}{|l|}}w_{0}^{\frac{2}{|l|}}H^{\frac{1}{|l|}}} \times \\ \frac{-(F_{th}|l|!\pi )^{\frac{1}{|l|}}w_{0}^{\frac{2}{|l|}}H^{\frac{1}{|l|}-1}}{2^{\frac{1}{|l|}}E_{0}^{\frac{1}{|l|}}|l|^{2}} \times \frac{2z\lambda^{2}}{\pi^{2}w_{0}^{4}}
\end{multline}
\begin{equation} \label{eq:39}
\frac{dD}{dz} = \frac{W_{-1}(G)}{1+W_{-1}(G)} \times \frac{H^{-1}}{|l|} \times \frac{2z\lambda^{2}}{\pi^{2}w_{0}^{4}}
\end{equation}
\begin{equation} \label{eq:40}
\frac{dD}{dz} = \frac{W_{-1}(G)}{1+W_{-1}(G)} \times \frac{2z\lambda^{2}}{|l|\pi^{2}w_{0}^{4}(\frac{z^{2}\lambda^{2}}{\pi^{2}w_{0}^{4}} + 1)}
\end{equation}
\begin{equation} \label{eq:41}
\frac{dD}{dz} = \frac{W_{-1}(G)}{1+W_{-1}(G)} \times \frac{2z\lambda^{2}}{|l|(z^{2}\lambda^{2} + \pi^{2}w_{0}^{4})}
\end{equation}
\\
From equations \ref{eq:16}, \ref{eq:20}, \ref{eq:22}, \ref{eq:35} and \ref{eq:41}:
\begin{equation} \label{eq:42}
\frac{dC}{dz} = \frac{\frac{dD}{dz}B - D\frac{dB}{dz}}{B^{2}} 
\end{equation}
\begin{multline} \label{eq:43}
	\frac{dC}{dz} = \left( \left( \frac{W_{-1}(G)}{1+W_{-1}(G)} \times \frac{2z\lambda^{2}}{|l|(z^{2}\lambda^{2} + \pi^{2}w_{0}^{4})} \times \right. \right. \\ \left. \left. \frac{-2}{|l|w_{0}^{2}(\frac{z^{2}\lambda^{2}}{\pi^{2}w_{0}^{4}} + 1)} \right) - \frac{W_{-1}(G)4z\lambda^{2}}{|l|I} \right) \div \\ \left(\frac{-2}{|l|w_{0}^{2}(\frac{z^{2}\lambda^{2}}{\pi^{2}w_{0}^{4}} + 1)}\right)^{2}
\end{multline}
\begin{multline} \label{eq:44}
	\frac{dC}{dz} = \left( \left( \frac{W_{-1}(G)}{1+W_{-1}(G)} \times \right. \right.\\ 
	\left. \left. \frac{-4z\lambda^{2}}{|l|^{2}w_{0}^{2}(\frac{z^{4}\lambda^{4}}{\pi^{2}w_{0}^{4}} + 2z^{2}\lambda^{2} + \pi^{2}w_{0}^{4})} \right) - \frac{W_{-1}(G)4z\lambda^{2}}{|l|I} \right) \div \\ \left(\frac{4}{|l|^{2}w_{0}^{4}(\frac{z^{4}\lambda^{4}}{\pi^{4}w_{0}^{8}} + \frac{2z^{2}\lambda^{2}}{\pi^{2}w_{0}^{4}} + 1)}\right)
\end{multline}
\begin{multline} \label{eq:45}
	\frac{dC}{dz} = \left( \left( \frac{W_{-1}(G)}{1+W_{-1}(G)} \times \frac{-4z\lambda^{2}}{|l|^{2}I} \right) - \frac{W_{-1}(G)4z\lambda^{2}}{|l|I} \right) \\ \div \left(\frac{4\pi^{2}w_{0}^{2}}{|l|^{2}I}\right)
\end{multline}
\begin{multline} \label{eq:46}
	\frac{dC}{dz} = \left( \frac{-W_{-1}(G)z\lambda^{2}}{(1+W_{-1}(G))|l|} - W_{-1}(G)z\lambda^{2} \right) \\ \times \frac{|l|}{\pi^{2}w_{0}^{2}}
\end{multline}
\begin{equation} \label{eq:47}
\frac{dC}{dz} = \frac{-z\lambda^{2}}{\pi^{2}w_{0}^{2}} \left( \frac{W_{-1}(G)}{(1+W_{-1}(G))} + W_{-1}(G)|l| \right)
\end{equation}
\\
From equations \ref{eq:21}, \ref{eq:29} and \ref{eq:47}:
\begin{equation} \label{eq:48}
\frac{d\rho}{dz} = \frac{d\rho}{dC} \times \frac{dC}{dz}
\end{equation}
\begin{equation} \label{eq:49}
\frac{d\rho}{dz} = \frac{1}{2\sqrt{C}} \times \frac{-z\lambda^{2}}{\pi^{2}w_{0}^{2}} \left( \frac{W_{-1}(G)}{(1+W_{-1}(G))} + W_{-1}(G)|l| \right)
\end{equation}
\\
To find the maxima we set $\frac{d\rho}{dz} = 0$ and rearrange for z, therefore:
\begin{equation} \label{eq:50}
0 = \frac{1}{(1+W_{-1}(G))} + |l|
\end{equation}
\begin{equation} \label{eq:51}
|l| + |l|W_{-1}(G)= -1
\end{equation}
\begin{equation} \label{eq:52}
W_{-1}(G)= \frac{-1-|l|}{|l|}
\end{equation}
\\
We now apply the definition of the Lambert Omega function once more, to carry out the reverse of the transform done previously (equations \ref{eq:11} and \ref{eq:12}):
\begin{equation} \label{eq:53}
G= \left(\frac{-1-|l|}{|l|}\right)e^{\frac{-1-|l|}{|l|}}
\end{equation}
Substituting in equation \ref{eq:18}:
\begin{equation} \label{eq:54}
\frac{-(F_{th}|l|!\pi )^{\frac{1}{|l|}}w_{0}^{\frac{2}{|l|}}\left(\frac{z^{2}\lambda^{2}}{\pi^{2}w_{0}^{4}} + 1\right)^{\frac{1}{|l|}}}{2^{\frac{1}{|l|}}E_{0}^{\frac{1}{|l|}}|l|} = \left(\frac{-1-|l|}{|l|}\right)e^{\frac{-1-|l|}{|l|}}
\end{equation}
\\
The $z$-value where the damage radius $\rho$ reaches its maximum is denoted $\chi$, where $z = \pm\chi$ for $\rho = \rho_{max}$, therefore:
\begin{equation} \label{eq:55}
\frac{-(F_{th}|l|!\pi )^{\frac{1}{|l|}}w_{0}^{\frac{2}{|l|}}\left(\frac{\chi^{2}\lambda^{2}}{\pi^{2}w_{0}^{4}} + 1\right)^{\frac{1}{|l|}}}{2^{\frac{1}{|l|}}E_{0}^{\frac{1}{|l|}}|l|} = \left(\frac{-1-|l|}{|l|}\right)e^{\frac{-1-|l|}{|l|}}
\end{equation}
\begin{equation} \label{eq:56}
\left(\frac{\chi^{2}\lambda^{2}}{\pi^{2}w_{0}^{4}} + 1\right)^{\frac{1}{|l|}} = \frac{(|l|+1)2^{\frac{1}{|l|}}E_{0}^{\frac{1}{|l|}}}{(F_{th}|l|!\pi )^{\frac{1}{|l|}}w_{0}^{\frac{2}{|l|}}}e^{\frac{-1-|l|}{|l|}}
\end{equation}
\begin{equation} \label{eq:57}
\frac{\chi^{2}\lambda^{2}}{\pi^{2}w_{0}^{4}} = \frac{(|l|+1)^{|l|}2E_{0}}{F_{th}|l|!\pi w_{0}^{2}}e^{-1-|l|} - 1
\end{equation}
\begin{equation} \label{eq:58}
\chi^{2} = \frac{(|l|+1)^{|l|}2\pi w_{0}^{2}E_{0}}{F_{th}|l|!\lambda^{2}}e^{-1-|l|} - \frac{\pi^{2}w_{0}^{4}}{\lambda^{2}}
\end{equation}
\begin{equation} \label{eq:59}
\chi = \sqrt{\frac{(|l|+1)^{|l|}2\pi w_{0}^{2}E_{0}}{F_{th}|l|!\lambda^{2}e^{|l|+1}} - \frac{\pi^{2}w_{0}^{4}}{\lambda^{2}}}
\end{equation}
\\
This expression has a similar form to the equivalent expression below for a Gaussian beam \cite{Samad06}, and simplifies to this for $l = 0$.
\begin{equation} \label{eq:60}
\chi = \sqrt{\frac{2\pi w_{0}^{2}E_{0}}{F_{th}\lambda^{2}e} - \frac{\pi^{2}w_{0}^{4}}{\lambda^{2}}}
\end{equation}

\section{Isolating the Damage Threshold}\label{sec:Threshold}

To isolate the damage threshold, we substitute $\chi$ (equation \ref{eq:59}) as the $z$-value into the expression for damage radius $\rho(z)$ (equation \ref{eq:145}). Values of $A$ and $B$ are replicated from equations \ref{eq:8} and \ref{eq:9}, with equation \ref{eq:3} substituted in for $w(z)$.
\begin{equation} \label{eq:62}
\rho (z) = \sqrt\frac{W_{-1}(AB)}{B}
\end{equation}
\begin{equation} \label{eq:63}
A = \left(\frac{F_{th}|l|!\pi \left(w_{0}\sqrt{1 + \frac{z^{2}\lambda^{2}}{\pi^{2} w_{0}^{4}}}\right)^{2(|l|+1)}}{2^{|l|+1}E_{0}}\right)^{\frac{1}{|l|}}
\end{equation}
\begin{equation} \label{eq:64}
B = \frac{-2}{|l|\left(w_{0}\sqrt{1 + \frac{z^{2}\lambda^{2}}{\pi^{2} w_{0}^{4}}}\right)^{2}}
\end{equation}
\\
When $z = \chi$, $\rho(z) = \rho_{max}$, therefore:
\begin{multline} \label{eq:65}
	\rho_{max}^{2} = W_{-1}\left(\left(\frac{F_{th}|l|!\pi \left(w_{0}\sqrt{1 + \frac{\chi^{2}\lambda^{2}}{\pi^{2} w_{0}^{4}}}\right)^{2(|l|+1)}}{2^{|l|+1}E_{0}}\right)^{\frac{1}{|l|}} \right. \\ \left. \times \frac{-2}{|l|\left(w_{0}\sqrt{1 + \frac{\chi^{2}\lambda^{2}}{\pi^{2} w_{0}^{4}}}\right)^{2}}\right) \div \frac{-2}{|l|\left(w_{0}\sqrt{1 + \frac{\chi^{2}\lambda^{2}}{\pi^{2} w_{0}^{4}}}\right)^{2}}
\end{multline}
\begin{multline} \label{eq:66}
	\rho_{max}^{2} = W_{-1}\left(\left(\frac{F_{th}|l|!\pi \left(w_{0}^{2} + \frac{\chi^{2}\lambda^{2}}{\pi^{2} w_{0}^{2}}\right)^{|l|+1}}{2^{|l|+1}E_{0}}\right)^{\frac{1}{|l|}} \right. \\ \left. \times \frac{-2}{|l|\left(w_{0}^{2} + \frac{\chi^{2}\lambda^{2}}{\pi^{2} w_{0}^{2}}\right)}\right) \div \frac{-2}{|l|\left(w_{0}^{2} + \frac{\chi^{2}\lambda^{2}}{\pi^{2} w_{0}^{2}}\right)}
\end{multline}
\begin{multline} \label{eq:67}
	\rho_{max}^{2} = W_{-1}\left(\frac{(F_{th}|l|!\pi)^{\frac{1}{|l|}} \left(w_{0}^{2} + \frac{\chi^{2}\lambda^{2}}{\pi^{2} w_{0}^{2}}\right)^{1+\frac{1}{|l|}}}{2^{1+\frac{1}{|l|}}E_{0}^{\frac{1}{|l|}}} \right. \\ \left. \times \frac{-2}{|l|\left(w_{0}^{2} + \frac{\chi^{2}\lambda^{2}}{\pi^{2} w_{0}^{2}}\right)}\right) \times \frac{|l|\left(w_{0}^{2} + \frac{\chi^{2}\lambda^{2}}{\pi^{2} w_{0}^{2}}\right)}{-2}
\end{multline}
\begin{multline} \label{eq:68}
	\frac{-2\rho_{max}^{2}}{|l|\left(w_{0}^{2} + \frac{\chi^{2}\lambda^{2}}{\pi^{2} w_{0}^{2}}\right)} = \\ W_{-1}\left(\frac{-(F_{th}|l|!\pi)^{\frac{1}{|l|}} \left(w_{0}^{2} + \frac{\chi^{2}\lambda^{2}}{\pi^{2} w_{0}^{2}}\right)^{\frac{1}{|l|}}}{2^{\frac{1}{|l|}}E_{0}^{\frac{1}{|l|}}|l|}\right)
\end{multline}
\\
We now need to substitute equation \ref{eq:59} into equation \ref{eq:68} in place of $\chi$. Considering the left hand side only:
\begin{equation} \label{eq:69}
LHS = \frac{-2\rho_{max}^{2}}{|l|\left(w_{0}^{2} + \frac{\chi^{2}\lambda^{2}}{\pi^{2} w_{0}^{2}}\right)}
\end{equation}
\begin{equation} \label{eq:70}
LHS = \frac{-2\rho_{max}^{2}}{|l|\left(w_{0}^{2} + \frac{\lambda^{2}}{\pi^{2} w_{0}^{2}}\left(\frac{(|l|+1)^{|l|}2\pi w_{0}^{2}E_{0}}{F_{th}|l|!\lambda^{2}e^{|l|+1}} - \frac{\pi^{2}w_{0}^{4}}{\lambda^{2}}\right)\right)}
\end{equation}
\begin{equation} \label{eq:71}
LHS = \frac{-2\rho_{max}^{2}}{|l|\left(w_{0}^{2} + \frac{(|l|+1)^{|l|}2E_{0}}{F_{th}|l|!\pi e^{|l|+1}} - w_{0}^{2}\right)}
\end{equation}
\begin{equation} \label{eq:72}
LHS = \frac{-2\rho_{max}^{2}F_{th}|l|!\pi e^{|l|+1}}{|l|(|l|+1)^{|l|}2E_{0}}
\end{equation}
\\
Now considering the right hand side:
\begin{equation} \label{eq:73}
RHS = W_{-1}\left(\frac{-(F_{th}|l|!\pi)^{\frac{1}{|l|}} \left(w_{0}^{2} + \frac{\chi^{2}\lambda^{2}}{\pi^{2} w_{0}^{2}}\right)^{\frac{1}{|l|}}}{2^{\frac{1}{|l|}}E_{0}^{\frac{1}{|l|}}|l|}\right)
\end{equation}
\begin{multline} \label{eq:74}
	RHS = W_{-1}\left(\frac{-(F_{th}|l|!\pi)^{\frac{1}{|l|}}}{2^{\frac{1}{|l|}}E_{0}^{\frac{1}{|l|}}|l|}\left(w_{0}^{2} + \frac{\lambda^{2}}{\pi^{2} w_{0}^{2}} \right. \right. \\ \left. \left. \times \left(\frac{(|l|+1)^{|l|}2\pi w_{0}^{2}E_{0}}{F_{th}|l|!\lambda^{2}e^{|l|+1}} - \frac{\pi^{2}w_{0}^{4}}{\lambda^{2}}\right)\right)^{\frac{1}{|l|}}\right)
\end{multline}
\begin{equation} \label{eq:75}
RHS = W_{-1}\left(\frac{-(F_{th}|l|!\pi)^{\frac{1}{|l|}}}{2^{\frac{1}{|l|}}E_{0}^{\frac{1}{|l|}}|l|}\left(\frac{(|l|+1)^{|l|}2E_{0}}{F_{th}|l|!\pi e^{|l|+1}}\right)^{\frac{1}{|l|}}\right)
\end{equation}
\begin{equation} \label{eq:76}
RHS = W_{-1}\left(\frac{-(F_{th}|l|!\pi)^{\frac{1}{|l|}}}{2^{\frac{1}{|l|}}E_{0}^{\frac{1}{|l|}}|l|} \times \frac{(|l|+1)2^{\frac{1}{|l|}}E_{0}^{\frac{1}{|l|}}}{(F_{th}|l|!\pi)^{\frac{1}{|l|}}e^{1+\frac{1}{|l|}}}\right)
\end{equation}
\begin{equation} \label{eq:77}
RHS = W_{-1}\left(\frac{|l|+1}{-|l|e^{1+\frac{1}{|l|}}}\right)
\end{equation}
\\
Setting $LHS = RHS$:
\begin{equation} \label{eq:78}
\frac{-2\rho_{max}^{2}F_{th}|l|!\pi e^{|l|+1}}{|l|(|l|+1)^{|l|}2E_{0}} = W_{-1}\left(\frac{|l|+1}{-|l|e^{1+\frac{1}{|l|}}}\right)
\end{equation}
\begin{equation} \label{eq:79}
F_{th} = \frac{-|l|(|l|+1)^{|l|}}{|l|!\pi e^{|l|+1}}W_{-1}\left(\frac{|l|+1}{-|l|e^{1+\frac{1}{|l|}}}\right)\frac{E_{0}}{\rho_{max}^{2}}
\end{equation}
\\
This equation gives the damage threshold as a function of the maximum damage radius $\rho_{max}$, for a given power $E_{0}$ and vortex charge $l$, allowing calculation of the ablation threshold from measurement of $\rho_{max}$.
\\
Simplifications of this formula for vortex beams of order $l = 1 - 5$ are shown below:
\begin{equation}\label{eq:79a}
\text{Vortex 1\textsuperscript{st} order ($l$ = 1, -1):}\hspace{0.4cm} F_{th} \approx 0.1723\frac{E_{0}}{\rho_{max}^{2}}
\end{equation}
\begin{equation}\label{eq:79b}
\text{Vortex 2\textsuperscript{nd} order ($l$ = 2, -2):}\hspace{0.4cm} F_{th} \approx 0.2139\frac{E_{0}}{\rho_{max}^{2}}
\end{equation}
\begin{equation}\label{eq:79c}
\text{Vortex 3\textsuperscript{rd} order ($l$ = 3, -3):}\hspace{0.4cm} F_{th} \approx 0.2487\frac{E_{0}}{\rho_{max}^{2}}
\end{equation}
\begin{equation}\label{eq:79d}
\text{Vortex 4\textsuperscript{th} order ($l$ = 4, -4):}\hspace{0.4cm} F_{th} \approx 0.2793\frac{E_{0}}{\rho_{max}^{2}}
\end{equation}
\begin{equation}\label{eq:79e}
\text{Vortex 5\textsuperscript{th} order ($l$ = 5, -5):}\hspace{0.4cm} F_{th} \approx 0.3068\frac{E_{0}}{\rho_{max}^{2}}
\end{equation}
\\
These have a similar form to the equivalent expression for a Gaussian beam \cite{Samad06}:
\\
\begin{equation}\label{eq:79f}
\text{Gaussian:}\hspace{0.4cm} F_{th} \approx 0.1171\frac{E_{0}}{\rho_{max}^{2}}
\end{equation}

\section{Limitations of the Lambert Omega Function} \label{sec:Omega}
The non-principal branch of the Lambert Omega function $W_{-1}(x)$ used to calculate $\rho_{out,max}$, is defined for real variables only when $0 \geq x \geq \frac{-1}{e}$. Therefore to find the limits (in the $z$-direction) of the solution defined in equation \ref{eq:145} we set:
\begin{equation} \label{eq:80}
0 \geq AB \geq \frac{-1}{e} 
\end{equation}
From the definitions of $A$ and $B$ in equations \ref{eq:8} and \ref{eq:9}, it is clear that the inequality $0 \geq AB$ is true, as $F_{th}$, $w(z)$, and $E_{0}$ are physical parameters with positive, real values.
Now we consider only the inequality $ AB \geq \frac{-1}{e}$. Substituting in equations \ref{eq:8} and \ref{eq:9} for $A$ and $B$ we get:
\begin{multline}\label{eq:81}
\left(\frac{F_{th}|l|!\pi \left(w_{0}\sqrt{1 + \frac{z^{2}\lambda^{2}}{\pi^{2} w_{0}^{4}}}\right)^{2(|l|+1)}}{2^{|l|+1}E_{0}}\right)^{\frac{1}{|l|}} \times \\ \frac{-2}{|l|\left(w_{0}\sqrt{1 + \frac{z^{2}\lambda^{2}}{\pi^{2} w_{0}^{4}}}\right)^{2}} \geq \frac{-1}{e}
\end{multline}
\begin{multline}\label{eq:82}
	\frac{(F_{th}|l|!\pi)^{\frac{1}{|l|}} \left(w_{0}^{2} + \frac{z^{2}\lambda^{2}}{\pi^{2} w_{0}^{2}}\right)^{{1+\frac{1}{|l|}}}}{2^{1+\frac{1}{|l|}}E_{0}^{\frac{1}{|l|}}} \\ \times \frac{-2}{|l|\left(w_{0}^{2} + \frac{z^{2}\lambda^{2}}{\pi^{2} w_{0}^{2}}\right)} \geq \frac{-1}{e} 
\end{multline}
\begin{equation} \label{eq:83}
	\frac{-(F_{th}|l|!\pi)^{\frac{1}{|l|}} \left(w_{0}^{2} + \frac{z^{2}\lambda^{2}}{\pi^{2} w_{0}^{2}}\right)^{\frac{1}{|l|}}}{|l|2^{\frac{1}{|l|}}E_{0}^{\frac{1}{|l|}}} \geq \frac{-1}{e} 
\end{equation}
\begin{equation} \label{eq:84}
	\frac{(F_{th}|l|!\pi)\left(w_{0}^{2} + \frac{z^{2}\lambda^{2}}{\pi^{2} w_{0}^{2}}\right)}{|l|^{|l|}2E_{0}} \leq \frac{1}{e^{|l|}} 
\end{equation}
\begin{equation} \label{eq:85}
	\frac{z^{2}\lambda^{2}}{\pi^{2} w_{0}^{2}} \leq \frac{|l|^{|l|}2E_{0}}{F_{th}|l|!\pi e^{|l|}} - w_{0}^{2} 
\end{equation}
\begin{equation} \label{eq:85a}
z^{2} \leq \frac{\pi^{2} w_{0}^{2}}{\lambda^{2}} \left(\frac{|l|^{|l|}2E_{0}}{F_{th}|l|!\pi e^{|l|}} - w_{0}^{2}\right) 
\end{equation}
Therefore:
\begin{equation} \label{eq:86}
z_{lim} = \frac{\pi w_{0}}{\lambda}\sqrt{\frac{|l|^{|l|}2E_{0}}{F_{th}|l|!\pi e^{|l|}} - w_{0}^{2}} 
\end{equation}
and
\begin{equation}\label{eq:86a}
-\left|z_{lim}\right| \leq z \leq \left|z_{lim}\right|
\end{equation}

This expression describes the minimum and maximum $z$-values that are defined using the Lambert Omega function. In order for equation \ref{eq:79} to be usable, we need to know when it is valid (i.e. when it is defined using the Lambert Omega function). For it to be valid, $\rho_{max}$ (which is reached when $z = \pm \chi$) must be defined, leading to the expression:
\begin{equation}\label{eq:87}
\chi \leq z_{lim}
\end{equation}
\\
Substituting in equations \ref{eq:59} and \ref{eq:86}:
\begin{equation}\label{eq:88}
\sqrt{\frac{(|l|+1)^{|l|}2\pi w_{0}^{2}E_{0}}{F_{th}|l|!\lambda^{2}e^{|l|+1}} - \frac{\pi^{2}w_{0}^{4}}{\lambda^{2}}} \leq \frac{\pi w_{0}}{\lambda}\sqrt{\frac{|l|^{|l|}2E_{0}}{F_{th}|l|!\pi e^{|l|}} - w_{0}^{2}}
\end{equation}
\begin{equation}\label{eq:89}
\frac{(|l|+1)^{|l|}2\pi w_{0}^{2}E_{0}}{F_{th}|l|!\lambda^{2}e^{|l|+1}} - \frac{\pi^{2}w_{0}^{4}}{\lambda^{2}} \leq \frac{|l|^{|l|}2\pi w_{0}^{2}E_{0}}{F_{th}|l|!\lambda^{2}e^{|l|}} -\frac{\pi^{2}w_{0}^{4}}{\lambda^{2}}
\end{equation}
\begin{equation}\label{eq:90}
\frac{(|l|+1)^{|l|}2\pi w_{0}^{2}E_{0}}{F_{th}|l|!\lambda^{2}e^{|l|+1}} \leq \frac{|l|^{|l|}2\pi w_{0}^{2}E_{0}}{F_{th}|l|!\lambda^{2}e^{|l|}}
\end{equation}
\begin{equation}\label{eq:91}
\frac{(|l|+1)^{|l|}}{e^{|l|+1}} \leq \frac{|l|^{|l|}}{e^{|l|}}
\end{equation}
\begin{equation}\label{eq:92}
(|l|+1)^{|l|} \leq |l|^{|l|}e
\end{equation}
Which is true for the case $l = 0$. For the nontrivial case where $l \neq 0$:
\begin{equation}\label{eq:93}
|l|\ln(|l|+1) \leq |l|\ln(|l|) + 1
\end{equation}
\begin{equation}\label{eq:94}
\ln(|l|+1) \leq \ln(|l|) + \frac{1}{|l|}
\end{equation}
\begin{equation}\label{eq:142}
0 \leq \ln(|l|) - \ln(|l|+1) + \frac{1}{|l|}
\end{equation}
\begin{figure}
	\centering
	\includegraphics[width=1\columnwidth]{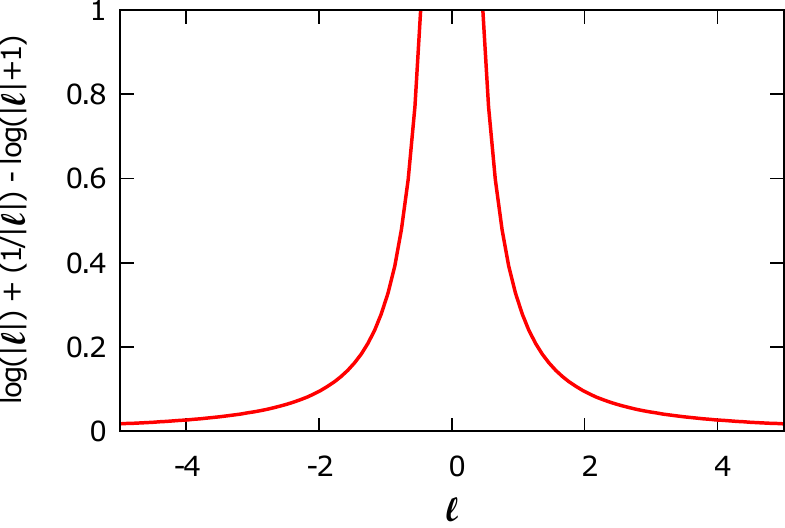}
	\textbf{\caption{\label{fig:Fig_3}Graph showing the behaviour of equation \ref{eq:142} for values of $l$ from $-5$ to $5$.}}
\end{figure}
We can evaluate equation \ref{eq:142} and see it is correct for small values of $l$ (Figure \ref{fig:Fig_3}). To verify this for larger values of $l$ we examine the limits of this expression as $l$ approaches $\infty$:
\begin{equation}\label{eq:143}
\lim_{l \to \infty } \left( \ln(|l|) - \ln(|l|+1) + \frac{1}{|l|} \right)
\end{equation}
We can separate this expression into two limits
\begin{equation}\label{eq:146}
\lim_{l \to \infty } \frac{1}{|l|} + \lim_{l \to \infty }\left( \ln(|l|) - \ln(|l|+1) \right)
\end{equation}
\begin{equation}\label{eq:147}
\lim_{l \to \infty } \frac{1}{|l|} = 0
\end{equation}
\begin{equation}\label{eq:148}
\lim_{l \to \infty }\left( \ln(|l|) - \ln(|l|+1) \right) = \lim_{l \to \infty }\left( \ln \left( \frac{|l|}{|l|+1} \right) \right)
\end{equation}
\begin{equation}\label{eq:149}
\lim_{l \to \infty }\left( \ln \left( \frac{|l|}{|l|+1} \right) \right) = 0
\end{equation}
Therefore we can see that
\begin{equation}\label{eq:150}
\lim_{l \to \infty } \left( \ln(|l|) - \ln(|l|+1) + \frac{1}{|l|} \right) = 0
\end{equation}
This indicates that equation \ref{eq:142} is true for $l \neq 0$. Therefore from equations \ref{eq:92} and \ref{eq:142} we can see that $\rho_{max}$ will always have a definite value according to equation \ref{eq:13} for any $l$.

\section{Pulse Superposition}\label{sec:PulseN}
\FloatBarrier

Having found an expression for the ablation threshold $F_{th}$, we would also like to be able to determine the number of pulses that correspond to this ablation threshold. For the Gaussian beam this has been found by defining pulse superposition $N$ at $z = \chi$ as the ratio of the sum of the intensity of all pulses hitting the sample to the intensity of a single pulse centred at $\chi$ \cite{Machado12}. Here we extend this definition to an optical vortex beam.
\\
\begin{figure}[h]
	\centering
	\includegraphics[width=1\columnwidth]{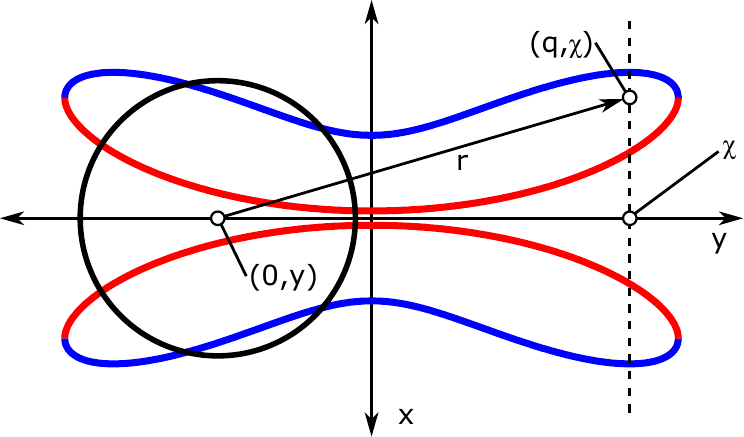}
	\textbf{\caption{\label{fig:Fig_4}Graph showing a diagonal scan for a vortex beam.}}
\end{figure}
\\
Consider the diagonal scan shown in Figure \ref{fig:Fig_4}, where a pulse hits the surface centred at $(0,y)$. The intensity created by this pulse at $(q,\chi)$ is given by:
\begin{equation} \label{eq:95}
F(q,y,z) = \frac{2^{|l|+1}(q^{2} + (\chi - y)^{2})^{|l|}E_{0}}{|l|!\pi w(z)^{2(|l|+1)}}e^{\frac{-2(q^{2} + (\chi - y)^{2})}{w(z)^{2}}}
\end{equation}
from equation \ref{eq:1}, as $r = \sqrt{q^{2} + (\chi - y)^{2}}$. At time $t = t_{0}$ a pulse hits the sample at $y = \chi$, generating the intensity $F(q,\chi,z)$. Here it generates the profile maximum $\rho_{max}$ and is therefore located at $z = \chi$ by definition. Therefore, $F(q,\chi,z) = F(q,\chi,\chi)$.
\\
The next pulse hits after some time $\frac{1}{f}$, (where $f$ is the repetition rate of the laser in Hz) at which point the sample has been displaced by $\frac{v_{y}}{f}$ and $\frac{v_{z}}{f}$ in the $y$ and $z$ directions respectively (where $v_{y}$ and $v_{z}$ are the translational speeds in the $y$ and $z$ directions respectively).
\\
The intensity at $(q,\chi)$ generated by the n$^{th}$ pulse is therefore given by:
\begin{equation} \label{eq:96}
F_{t_{0} + \frac{n}{f}} = \frac{2^{|l|+1}(q^{2} + (nv_{y}/f)^{2})^{|l|}E_{0}}{|l|!\pi w(\chi + (nv_{z}/f))^{2(|l|+1)}}e^{\frac{-2(q^{2} + (nv_{y}/f)^{2})}{w(\chi + (nv_{z}/f))^{2}}}
\end{equation}
\\
The total intensity accumulated at $(q,\chi)$, $F_{tot}$ is the sum of all pulses that hit the sample:
\begin{equation} \label{eq:97}
F_{tot} = \sum_{n=-\infty}^{\infty} F_{t_{0} + \frac{n}{f}}
\end{equation}
\begin{multline} \label{eq:98}
F_{tot} = \sum_{n=-\infty}^{\infty} \frac{2^{|l|+1}(q^{2} + (nv_{y}/f)^{2})^{|l|}E_{0}}{|l|!\pi w(\chi + (nv_{z}/f))^{2(|l|+1)}}\\ \times e^{\frac{-2(q^{2} + (nv_{y}/f)^{2})}{w(\chi + (nv_{z}/f))^{2}}}
\end{multline}
If we assume that the spot size $w$ does not change significantly around $\chi$ and can be considered $= w(\chi)$, then:
\begin{multline} \label{eq:99}
F_{tot} = \frac{2^{|l|+1}E_{0}}{|l|!\pi} \times \\ \sum_{n=-\infty}^{\infty} \frac{(q^{2} + (nv_{y}/f)^{2})^{|l|}}{w(\chi)^{2(|l|+1)}}e^{\frac{-2(q^{2} + (nv_{y}/f)^{2})}{w(\chi)^{2}}}
\end{multline}
From equation \ref{eq:59}:
\begin{equation} \label{eq:100}
\chi = \sqrt{\frac{(|l|+1)^{|l|}2\pi w_{0}^{2}E_{0}}{F_{th}|l|!\lambda^{2}e^{|l|+1}} - \frac{\pi^{2}w_{0}^{4}}{\lambda^{2}}}
\end{equation}
Substituting this into the expression for beam waist (equation \ref{eq:3}) we obtain:
\begin{equation} \label{eq:101}
w(\chi) = w_{0}\sqrt{1 + \left(\frac{(|l|+1)^{|l|}2\pi w_{0}^{2}E_{0}}{F_{th}|l|!\lambda^{2}e^{|l|+1}} - \frac{\pi^{2}w_{0}^{4}}{\lambda^{2}}\right)\frac{\lambda^{2}}{\pi^{2}w_0^{4}}}
\end{equation}
\begin{equation} \label{eq:102}
w(\chi) = \sqrt{\frac{(|l|+1)^{|l|}2E_{0}}{F_{th}|l|!\pi e^{|l|+1}}}
\end{equation}
From equation \ref{eq:79} we know:
\begin{equation} \label{eq:103}
F_{th} = \frac{-|l|(|l|+1)^{|l|}E_{0}}{\rho_{max}^{2}|l|!\pi e^{|l|+1}}W_{-1}\left(\frac{|l|+1}{-|l|e^{1+\frac{1}{|l|}}}\right)
\end{equation}
Substituting equation \ref{eq:79} into equation \ref{eq:102}:
\begin{equation} \label{eq:104}
w(\chi) = \sqrt{\frac{2\rho_{max}^{2}}{-|l|}\left(W_{-1}\left(\frac{|l|+1}{-|l|e^{1+\frac{1}{|l|}}}\right)\right)^{-1}}
\end{equation}
Let:
\begin{equation} \label{eq:105}
L = \frac{|l|+1}{-|l|e^{1+\frac{1}{|l|}}}
\end{equation}
Such that:
\begin{equation} \label{eq:106}
w(\chi) = \sqrt{\frac{2\rho_{max}^{2}}{-|l|W_{-1}(L)}}
\end{equation}
Substituting equation \ref{eq:106} into equation \ref{eq:99}
\begin{multline} \label{eq:107}
F_{tot} = \frac{2^{|l|+1}E_{0}}{|l|!\pi} \sum_{n=-\infty}^{\infty} \frac{(q^{2} + (nv_{y}/f)^{2})^{|l|}}{(\frac{2\rho_{max}^{2}}{-|l|W_{-1}(L)})^{|l|+1}} \times \\e^{\frac{-2(q^{2} + (nv_{y}/f)^{2})}{2\rho_{max}^{2} / (-|l|W_{-1}(L))}}
\end{multline}
\begin{multline} \label{eq:108}
F_{tot} = \frac{(-|l|W_{-1}(L))^{|l|+1}E_{0}}{|l|!\pi \rho_{max}^{2(|l|+1)}} \times \\ \sum_{n=-\infty}^{\infty} \left(q^{2} + \left(\frac{nv_{y}}{f}\right)^{2}\right)^{|l|}e^{\frac{|l|W_{-1}(L)(q^{2} + (nv_{y}/f)^{2})}{\rho_{max}^{2}}}
\end{multline}
\begin{multline} \label{eq:109}
F_{tot} = \frac{(-|l|W_{-1}(L))^{|l|+1}E_{0}}{|l|!\pi \rho_{max}^{2(|l|+1)}}e^{\frac{|l|W_{-1}(L)q^{2}}{\rho_{max}^{2}}} \times \\ \sum_{n=-\infty}^{\infty} \left(q^{2} + \left(\frac{nv_{y}}{f}\right)^{2}\right)^{|l|}e^{\frac{|l|W_{-1}(L)(nv_{y}/f)^{2}}{\rho_{max}^{2}}}
\end{multline}
To obtain pulse superposition $N$, we normalise $F_{tot}$ to the intensity $F_{0}$ of a single pulse centred at $\chi$. First we must evaluate $F_{0}$ as $F_{t_{0} + \frac{n}{f}}$ when $n = 0$. Using equation \ref{eq:96}:
\begin{equation} \label{eq:110}
F_{0} = \frac{2^{|l|+1}q^{2|l|}E_{0}}{|l|!\pi w(\chi)^{2(|l|+1)}}e^{\frac{-2q^{2}}{w(\chi)^{2}}}
\end{equation}
Substituting equation \ref{eq:106} into equation \ref{eq:110}:
\begin{equation} \label{eq:111}
F_{0} = \frac{2^{|l|+1}q^{2|l|}E_{0}}{|l|!\pi \left(\frac{2\rho_{max}^{2}}{-|l|W_{-1}(L)}\right)^{(|l|+1)}}e^{\frac{-2q^{2}}{\left(\frac{2\rho_{max}^{2}}{-|l|W_{-1}(L)}\right)}}
\end{equation}
\begin{equation} \label{eq:112}
F_{0} = \frac{(-|l|W_{-1}(L))^{|l|+1}q^{2|l|}E_{0}}{|l|!\pi \rho_{max}^{2(|l|+1)}}e^{\frac{q^{2}|l|W_{-1}(L)}{\rho_{max}^{2}}}
\end{equation}
Now setting $N = F_{tot} / F_{0}$, and substituting in equations \ref{eq:109} and \ref{eq:112}
\begin{equation} \label{eq:113}
N = \frac{1}{q^{2|l|}}\sum_{n=-\infty}^{\infty} \left(q^{2} + \left(\frac{nv_{y}}{f}\right)^{2}\right)^{|l|}e^{|l|W_{-1}(L)\left(\frac{nv_{y}}{f\rho_{max}}\right)^{2}}
\end{equation}
\begin{equation} \label{eq:114}
N = \sum_{n=-\infty}^{\infty} \left(1 + \left(\frac{nv_{y}}{fq}\right)^{2}\right)^{|l|}e^{|l|W_{-1}(L)\left(\frac{nv_{y}}{f\rho_{max}}\right)^{2}}
\end{equation}
Unlike for a Gaussian beam, N clearly varies along the x-axis (length q). $F_{th}$ is calculated using the position $\rho_{max}$ along the x-axis, so to calculate the corresponding N we set $q = \rho_{max}$:
\begin{equation} \label{eq:115}
N = \sum_{n=-\infty}^{\infty} \left(1 + n^{2}\left(\frac{v_{y}}{f\rho_{max}}\right)^{2}\right)^{|l|}e^{|l|W_{-1}(L)\left(\frac{nv_{y}}{f\rho_{max}}\right)^{2}}
\end{equation}
We can define:
\begin{equation} \label{eq:116}
H^{2} = -W_{-1}(L)
\end{equation}
where $H \in \Re$, as $W_{-1}(L) < 0$ for any $L \in \Re$ by definition \cite{Corless96}. We can also define:
\begin{equation} \label{eq:117}
K = \frac{v_{y}}{f\rho_{max}}
\end{equation}
where $K \in \Re$, as $v_{y}$, $f$ and $\rho_{max}$ are all physical parameters with real positive values.
Therefore:
\begin{equation} \label{eq:118}
N = \sum_{n=-\infty}^{\infty} \left(1 + K^{2}n^{2}\right)^{|l|}e^{-|l|H^{2}K^{2}n^{2}}
\end{equation}
Unlike for a Gaussian beam, this expression does not have an analytical solution, therefore must be summed numerically. For this to result in a finite $N$, we must check the convergence of this summation.
Firstly, we can write $N$ as the sum of two sums (from equation \ref{eq:118}):
\begin{equation} \label{eq:138}
N = S_{1} + S_{2}
\end{equation}
\begin{equation} \label{eq:139}
S_{1} = \sum_{n=0}^{\infty} \left(1 + K^{2}n^{2}\right)^{|l|}e^{-|l|H^{2}K^{2}n^{2}}
\end{equation}
\begin{equation} \label{eq:140}
S_{2} = \sum_{n=-\infty}^{0} \left(1 + K^{2}n^{2}\right)^{|l|}e^{-|l|H^{2}K^{2}n^{2}}
\end{equation}
Comparing these pointwise, it is clear that $S_{1} = S_{2}$, therefore we can write:
\begin{equation} \label{eq:141}
N = 2\sum_{n=0}^{\infty} \left(1 + K^{2}n^{2}\right)^{|l|}e^{-|l|H^{2}K^{2}n^{2}}
\end{equation}
Now, let:
\begin{equation} \label{eq:119}
M = 2 max(K^{2}, 1)
\end{equation}
Therefore:
\begin{equation} \label{eq:120}
1 + K^{2}n^{2} \leq M + \frac{Mn^{2}}{2}
\end{equation}
The right hand side of equation \ref{eq:120} resembles the first few terms of the Taylor Expansion for $e^{n}$:
\begin{equation} \label{eq:121}
e^{n} = \sum_{k=0}^{\infty} \frac{n^{k}}{k!} = 1 + n + \frac{n^{2}}{2!} + \frac{n^{3}}{3!} + ......
\end{equation}
Therefore we can also say:
\begin{equation} \label{eq:122}
1 + K^{2}n^{2} \leq M + \frac{Mn^{2}}{2} \leq Me^{n}
\end{equation}
We can therefore set upper and lower bounds on $N$:
\begin{equation} \label{eq:123}
0 \leq N \leq N_{u} = 2\sum_{n=0}^{\infty} \left(Me^{n}\right)^{|l|}e^{-|l|H^{2}K^{2}n^{2}}
\end{equation}
\begin{equation} \label{eq:124}
0 \leq N \leq N_{u} = 2\sum_{n=0}^{\infty} M^{|l|}e^{|l|(n - H^{2}K^{2}n^{2})}
\end{equation}
We can see that the expression for the upper bound $N_{u}$ satisfies the conditions for the integral test (namely that it is positive and decreasing), and is therefore convergent, as $H \in \Re$ and $K \in \Re$. Therefore, since $N_{u} > N \geq 0$, $N$ is also convergent according to the limit comparison test \cite{Bartle00}. 
\\
Because $N$ has a finite sum, we can say that for every $\varepsilon > 0$, there exists a natural number $m$ such that for all $n \geq m$, the terms $N_{n}$ satisfy $|N_{n} - N| < \varepsilon$ \cite{Bartle00}. Therefore:
\begin{equation} \label{eq:125}
N = \sum_{n=-m}^{m} \left(1 + K^{2}n^{2}\right)^{|l|}e^{-|l|H^{2}K^{2}n^{2}}
\end{equation}
where:
\begin{equation} \label{eq:126}
K = \frac{v_{y}}{f\rho_{max}}
\end{equation}
\begin{equation} \label{eq:127}
H^{2} = -W_{-1}(L) = -W_{-1}\left(\frac{|l|+1}{-|l|e^{1+\frac{1}{|l|}}}\right)
\end{equation}
We can again invoke the symmetry of $N$ around $n = 0$ to write:
\begin{equation} \label{eq:137}
N = 2\sum_{n=0}^{m} \left(1 + K^{2}n^{2}\right)^{|l|}e^{-|l|H^{2}K^{2}n^{2}}
\end{equation}
We can then find an expression for $m$ in terms of $\varepsilon$:
\begin{equation} \label{eq:128}
\varepsilon = \left(1 + K^{2}m^{2}\right)^{|l|}e^{-|l|H^{2}K^{2}m^{2}}
\end{equation}
\begin{equation} \label{eq:129}
\varepsilon^{\frac{1}{|l|}} = \left(1 + K^{2}m^{2}\right)e^{-H^{2}K^{2}m^{2}}
\end{equation}
\begin{equation} \label{eq:130}
-H^{2}\varepsilon^{\frac{1}{|l|}} = -H^{2}\left(1 + K^{2}m^{2}\right)e^{-H^{2}K^{2}m^{2}}
\end{equation}
\begin{multline} \label{eq:131}
-H^{2}\varepsilon^{\frac{1}{|l|}}e^{-H^{2}} = -H^{2}\left(1 + K^{2}m^{2}\right)\\e^{-H^{2}K^{2}m^{2}}e^{-H^{2}}
\end{multline}
\begin{multline} \label{eq:132}
-H^{2}\varepsilon^{\frac{1}{|l|}}e^{-H^{2}} = -H^{2}\left(1 + K^{2}m^{2}\right)\\e^{-H^{2}(1 + K^{2}m^{2})}
\end{multline}
We can now use the Lambert Omega function again to rearrange this:
\begin{equation} \label{eq:133}
-H^{2}(1 + K^{2}m^{2}) = W_{-1}(-H^{2}\varepsilon^{\frac{1}{|l|}}e^{-H^{2}})
\end{equation}
And rearranging for $\varepsilon$:
\begin{equation} \label{eq:134}
1 + K^{2}m^{2} = \frac{W_{-1}(-H^{2}\varepsilon^{\frac{1}{|l|}}e^{-H^{2}})}{-H^{2}}
\end{equation}
\begin{equation} \label{eq:135}
m^{2} = \frac{W_{-1}(-H^{2}\varepsilon^{\frac{1}{|l|}}e^{-H^{2}})}{-H^{2}K^{2}} - \frac{1}{K^{2}}
\end{equation}
\begin{equation} \label{eq:136}
m = \sqrt{\frac{W_{-1}(-H^{2}\varepsilon^{\frac{1}{|l|}}e^{-H^{2}})}{-H^{2}K^{2}} - \frac{1}{K^{2}}}
\end{equation}
This allows us to calculate $N$ to arbitrary precision by choosing an arbitrarily small $\varepsilon$, calculating $m$ using equation \ref{eq:136}, and evaluating equation \ref{eq:137} numerically.

\section{Conclusions}
We have presented a formalism of the diagonal scan method for determining the ablation threshold and pusle superposition for femtosecond vortex pulses. To the author's knowledge this is the first ablation threshold technique developed for pulses with non-Gaussian spatial profiles.
\\
The method allows calculation of the ablation threshold from the measurement of a single feature, allowing investigations of ablation threshold and incubation effects to be carried out promptly. The extension of the method to non-Gaussian beams will allow new avenues of research to be explored, enabling characterisation of the ablation threshold and incubation behaviour for a material with femtosecond vortex pulses.

\section{Acknowledgements}
The authors acknowledge financial support from the New Zealand Ministry of Business, Innovation and Employment Grants UOAX1202 (Laser Microfabrication and Micromachining) and UOAX1416 (Tailored Beam Shapes for Fast, Efficient and Precise Femtosecond Laser Micromachining), and would like to thank Mr. Andy (Xindi) Wang for assistance with real analysis calculations.

\bibliographystyle{ieeetr} 
\bibliography{References}

\end{document}